\documentclass[aps,prb,twocolumn,longbibliography,superscriptaddress]{revtex4-1}
\usepackage{amsmath,amssymb,graphicx}
\usepackage{makecell}
\usepackage{bm}
\usepackage{color}
\usepackage[colorlinks]{hyperref}
\hypersetup{
	linkcolor = {blue},
    citecolor = {blue},
    urlcolor = {blue}
}
\definecolor{Nathanblue}{rgb}{0.96,0.24,0.00}

\begin{document}

\title{Topological quantum matter in synthetic dimensions}

\date{\today}

\author{Tomoki Ozawa}
\affiliation{Interdisciplinary Theoretical and Mathematical Sciences Program (iTHEMS), RIKEN, Wako, Saitama 351-0198, Japan}
\author{Hannah M. Price}
\affiliation{School of Physics and Astronomy, University of Birmingham, Edgbaston, Birmingham B15 2TT, United Kingdom}

\begin{abstract}
In the field of quantum simulation of condensed matter phenomena by artificially engineering the Hamiltonian of an atomic, molecular or optical system, the concept of `synthetic dimensions' has recently emerged as a powerful way to emulate phenomena such as topological phases of matter, which are now of great interest across many areas of physics. The main idea of a synthetic dimension is to couple together suitable degrees of freedom, such as a set of internal atomic states, in order to mimic the motion of a particle along an extra spatial dimension. This approach provides a way to engineer lattice Hamiltonians and enables the realisation of higher-dimensional topological models in platforms with lower dimensionality. We give an overview of the recent progress in studying topological matter in synthetic dimensions. After reviewing proposals and realizations in various  setups, we discuss future prospects in many-body physics, applications, and topological effects in three or more spatial dimensions.
\end{abstract}

\maketitle
\section{Introduction}
Topological quantum matter is a rapidly-growing field, in which topological concepts are exploited to discover and classify new phases of matter~\cite{Hasan:2010RMP,Qi:2011RMP,Chiu:2016RMP}. Unusually, these states fall outside of the conventional Landau paradigm in which phases of matter are characterised by local order parameters. Instead, topological quantum states are associated with global invariants that are constrained to take integer values and so cannot be continuously changed under small perturbations of the system. This property underlies fascinating phenomena in topological matter, such as the existence of robust one-way edge modes with potential applications in future devices. 

Although topological quantum states were first discovered in solid-state materials~\cite{Klitzing:1980PRL}, in the last decade they have been  engineered across a wide-range of systems, including photonics~\cite{Lu:2014NatPhot,lu2016topological, khanikaev2017two, ozawa2018topological} and ultracold atomic gases~\cite{Goldman:2014ROPP, Goldman:2016NatPhys, cooper2018topological}. These systems provide opportunities to go beyond what is possible in real materials, taking advantage of the high controllability and flexibility of these platforms. In this direction, the development of `synthetic dimensions'  provides a powerful approach for pushing forward the exploration of topological physics with atoms and photons.  

The central idea of `synthetic dimensions' is to exploit and harness some of the internal degrees of freedom of atoms or photons, so as to simulate motion along extra spatial directions~\cite{Boada:2012PRL,Celi:2014PRL}. In effect, this enables a lower-dimensional system to effectively simulate the behaviour of a higher-dimensional system. For example, a system in $D$ real spatial dimensions can mimic a system with $(D+d)$ effective spatial dimensions, if $d$ synthetic dimensions are added. This raises the intriguing possibility of probing systems with effectively more than three spatial dimensions in the laboratory.

In this Perspective, our aim is to present the exciting recent progress on topological quantum matter in synthetic dimensions. We begin by briefly discussing spatial dimensionality and the simulation of topological matter; this discussion is followed by an introduction of synthetic dimensions in ultracold gases and then in photonics. As an outlook, we review perspectives for future research, including prospects for exploring exotic many-body physics, applications and higher spatial dimensions. \\

\section{Spatial dimensionality}

\noindent At a basic level, the number of spatial dimensions corresponds to the number of independent directions along which an object can move. From this point of view, the spatial dimensionality of a system can be engineered by adding or taking away such independent directions. On the one hand, using a higher-dimensional system to explore lower spatial dimensions is very common within physics; for example, optical potentials are routinely applied to cold atom set-ups to effectively restrict atomic motion to only one or two spatial dimensions so as to study 1D and 2D physical phenomena~\cite{bloch2008many}. 

On the other hand, using a lower-dimensional system to explore higher-dimensional physics can be harder to visualise, as the extra dimensions are not `real' spatial dimensions but arise through a re-interpretation of the underlying physics. In this direction, several different conceptual approaches have been developed over recent decades. Perhaps the simplest, theoretically, is to directly embed a higher-dimensional lattice within a lower-dimensional space, for example, by arranging all sites of a 2D lattice along a 1D line and then adding the appropriate connectivity between sites as proposed with qubits~\cite{Tsomokos:2010PRA} and photonic lattices~\cite{Jukic:2013PRA}. However, this method is experimentally challenging as it requires precise engineering of both near- and long-range hoppings, and so it has not yet been used to explore topological systems. A second approach, which has been implemented experimentally, harnesses a mathematical mapping between certain higher-dimensional static lattices and lower-dimensional time-dependent systems, called `topological pumps'~\cite{Thouless:1983,Kraus:2012a,Kraus:2012b,Verbin:2013,Kraus:2013,Verbin:2015,Lohse:2016,Nakajima:2016}. Although the full dynamics is not captured by the mapping, as it replaces some dynamical variables with external parameters, certain important signatures of the higher-dimensional topology can manifest in topological pumps, as shown in recent photonic~\cite{Kraus:2012a,Kraus:2012b,Verbin:2013,Kraus:2013,Verbin:2015} and cold atomic~\cite{Lohse:2016,Nakajima:2016} experiments. 

\begin{figure*}
%\resizebox{0.85\textwidth}{!}{\includegraphics{intropicture.png}}
\resizebox{0.85\textwidth}{!}{\includegraphics{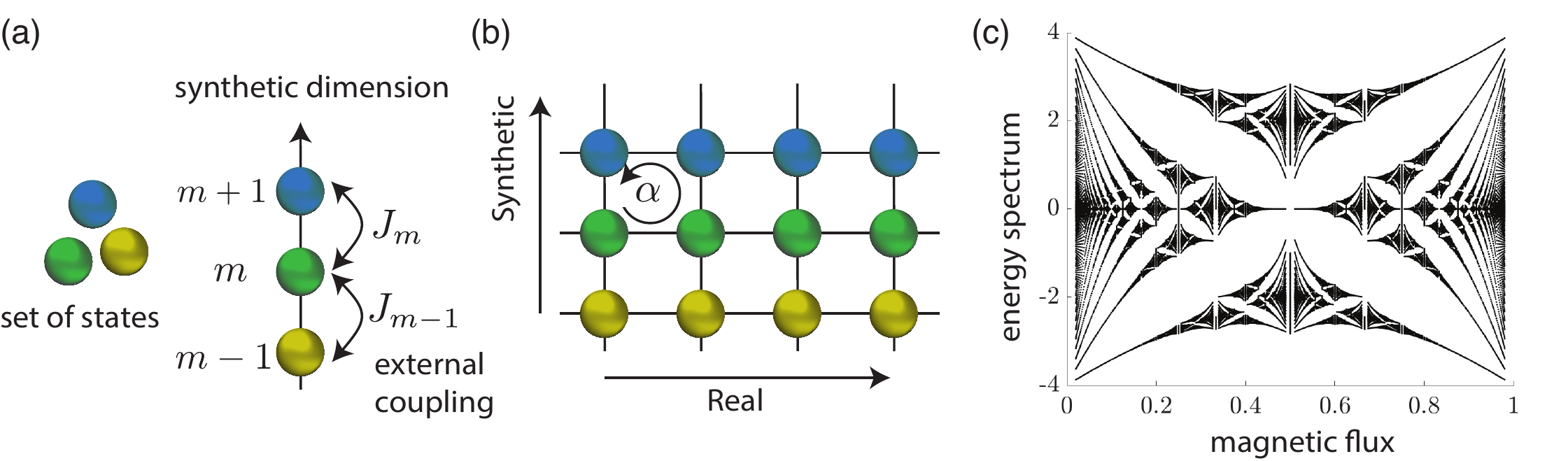}}
\caption{{\bf The concept of synthetic dimension.} (a) A set of states is re-interpreted as a set of lattice sites along a synthetic dimension, indexed here by $m$ and with `hopping amplitudes', $J_m$, from external couplings. (b) When motion along real spatial dimensions (or extra synthetic dimensions) is included, this can be used to realise topological systems, such as the Harper-Hofstadter model (Eq.~\ref{eq:HH}), in which each  plaquette is pierced by $\alpha$ magnetic flux quanta. (c) Even for narrow systems, the Harper-Hofstadter model retains key signatures~\cite{Celi:2014PRL}, such as hallmarks of the fractal energy spectrum, known as the Hofstadter butterfly, as shown here for periodic boundary conditions and only three sites in one direction. The persistence of such behaviour in narrow systems is advantageous for synthetic dimension experiments where the number of synthetic sites is often limited.} \label{fig:syntheticdimension}
\end{figure*}

Synthetic dimensions provide another way for a lower-dimensional set-up to explore higher-dimensional systems, which is experimentally fruitful and captures the full higher-dimensional dynamics. The main concept behind this approach is to identify a set of initially uncoupled degrees of freedom for a particle, and to reinterpret these as lattice sites along a spatial dimension~\cite{Boada:2012PRL}, as shown schematically in Fig.~\ref{fig:syntheticdimension}(a). Adding suitable external couplings within this set of degrees of freedom allows a particle to `hop' along this extra dimension, just as it would hop between sites in a lattice. The research in the area of synthetic dimensions has taken off and has opened the way for the simulation of topological states of matter~\cite{Celi:2014PRL}. \\

\section{Topological matter}

\noindent Depending on the spatial dimensionality, there is a wealth of different possible topological phases of matter, including topological insulators, topological superconductors and the quantum Hall phases~\cite{Hasan:2010RMP,Qi:2011RMP,Chiu:2016RMP}. The latter were in fact the first topological states to be discovered, when 2D electronic systems were subjected to strong magnetic fields in the early 1980s (Ref.~\onlinecite{Klitzing:1980PRL}). In these systems, the electronic energy bands are labelled by non-zero topological integers called Chern numbers~\cite{Thouless:1982PRL}, leading to a precise quantisation of the Hall conductance. 

Over the past decade, there has been great interest in realising topological states of matter -- such as quantum Hall phases -- outside of traditional condensed matter materials. However, in platforms such as cold atomic gases and photonics, the particles are neutral, and so do not respond like electrons to gauge fields. This has lead to the development of artificial gauge fields, whereby these systems are made to mimic electronic materials~\cite{lu2016topological, khanikaev2017two, ozawa2018topological,Goldman:2014ROPP, Goldman:2016NatPhys, cooper2018topological}. For particles hopping in a deep lattice, this requires experimental techniques that reproduce the tight-binding Peierls substitution, which is to capture the effects of a magnetic vector potential, ${\bf A} (\mathbf{r})$, on a charged particle as: 
\begin{eqnarray}
-   J_{{\bf r}} \hat{a}^\dagger_{{\bf r}+\Delta \mathbf{r} } \hat{a}_{\bf r} \rightarrow  -  J_{\bf r} \hat{a}
^\dagger_{{\bf r}+\Delta \mathbf{r} } \left( e^{i \int_{{\bf r}}^{{\bf r} + \Delta \mathbf{r} } {\bf A} ({\bf r}') \cdot \text{d} {\bf r}'}\right)
 \hat{a}_{\bf r} ,\qquad \label{eq:peierls}
\end{eqnarray}
where $\hat{a}_{\bf r}^\dagger$ ($\hat{a}_{\bf r}$) creates (destroys) a particle at a lattice site at position ${\bf r}$, with $J_r$ being the tunnelling amplitudes and $\Delta \mathbf{r} $ being the lattice vector connecting the sites. Extending this to a 2D square lattice in a uniform magnetic field, the Peierls substitution leads to the famous Harper-Hofstadter Hamiltonian~\cite{Harper:1955PPSA,Hofstadter:1976PRB}: 
\begin{eqnarray}
H = -  \sum_{x, y} J_{x, y} \left[ \hat{a}^\dagger_{x+1,y} \hat{a}_{x,y} + \hat{a}^\dagger_{x,y+1} \hat{a}_{x,y} e^{i \phi (x)} + \text{h.c.} \right] , \qquad \label{eq:HH}
\end{eqnarray}
which is a seminal quantum Hall model, here written in the Landau gauge, with $\phi (x) = 2\pi \alpha x$ and $\alpha$ being the number of magnetic flux quanta per unit cell. As can be seen, the required phase, $\phi$, has a simple spatial dependence that corresponds, around a closed loop in the lattice, to the Aharanov-Bohm phase of a charged particle. To date, various techniques have been developed to realise this physics for neutral particles~\cite{lu2016topological, khanikaev2017two, ozawa2018topological,Goldman:2014ROPP, Goldman:2016NatPhys, cooper2018topological}, but a key advantage of synthetic dimensions is that such phases can be engineered very naturally in the external couplings along the extra dimension. Importantly for experiments, many properties of such topological models are also robust in many cases~\cite{Celi:2014PRL} against issues such as inhomogeneous hopping amplitudes, $J_{x,y}$, and narrowing of the system widths as shown in (Fig.~\ref{fig:syntheticdimension}) which  can  be  limitations  of  certain synthetic dimension realisations, as discussed below. \\

\section{Synthetic dimensions in atomic gases} 

\noindent Historically, some early works similar to what we now call synthetic dimensions included the exploration of dimensional deconstruction in high-energy physics~\cite{ArkaniHamed:2001PRL}, where the gauge groups present in a theory were viewed as a synthetic dimension at low energy. Other works developed `kicked rotor' models~\cite{Casati:PRL1989,Edge:PRL2012}, where incommensurate frequencies were introduced to mimic extra dimensions, with experimental implementations in cold atoms~\cite{Casati:PRL1989,Moore:1995PRL,Manai:2015PRL,Chabe:2008PRL}. The more recent interest in synthetic dimensions began in 2014, thanks to a proposal~\cite{Celi:2014PRL} for engineering a topological system using a synthetic dimension of atomic states~\cite{Boada:2012PRL}. This proposal was rapidly followed by two independent experimental implementations~\cite{Mancini:2015Science,Stuhl:2015Science}, sparking off many different research directions in atomic gases, ranging from further developments of atomic-state schemes~\cite{Livi:2016PRL,Kolkowitz:2017Nature}, to experiments with momentum states~\cite{Gadway:2015PRA,Meier:2016PRA,Meier:2016NatComm,An:2017SciAdv,An:2018PRL,Viebahn:2019} and other studies harnessing harmonic oscillator~\cite{Price:2017PRA}, molecular~\cite{Sundar:2018SciRep} and Floquet~\cite{Martin:2017PRX,Baum:2018PRL,Peng:2018PRB,Andrijauskas:2018NJP} modes. We briefly compare these schemes at the non-interacting single-particle level, highlighting the most important characteristics in Table 1. \\

\begin{table*}[htb]
  \caption{Summary of various proposals and realizations for synthetic dimensions}
  \begin{tabular}{|l|l|l|r|} \hline
    Physical system & What states are used & How states are coupled & \thead{Number of \\ extra dimensions studied} \\ \hline \hline
    Atoms~\cite{Celi:2014PRL,Mancini:2015Science,Stuhl:2015Science} & Hyperfine states & Raman lasers & 1 (only few sites) \\ \hline
    Atoms~\cite{Livi:2016PRL,Kolkowitz:2017Nature} & Electronic states & Clock lasers & 1 (only two sites) \\ \hline
    Atoms~\cite{Sundar:2018SciRep} & Angular momentum of molecules & Microwaves & 1 - 2 \\ \hline
    Atoms~\cite{Gadway:2015PRA,Meier:2016PRA} & Momentum states & Bragg transitions &  No upper limit \\ \hline
    Atoms/Photons~\cite{Price:2017PRA,Lustig:2018arXiv} & Spatial eigenmodes & Shaking & 1 - 3 \\ \hline
    Photons~\cite{Luo:2015NatComm,Cardano:2017NatComm,Wang:2018PRL}  & Orbital angular momentum & Spatial light modulator& 1 \\ \hline
    Photons~\cite{Ozawa:2016PRA,Yuan:2016OptLett} & Frequency modes & Temporal modulation & No upper limit \\ \hline
    Photons~\cite{Ozawa:2017PRL} & Angular coordinate of resonator & Dispersion of resonator & 1 (continuous dimension) \\ \hline
    Photons~\cite{Schreiber:2010PRL,Regensburger:2011PRL,Wimmer:2017NatPhys,Chen:2018PRL} &  Arrival time of pulses & Coupled optical paths of different lengths & 1 - 2 \\ \hline
    Generic~\cite{Martin:2017PRX,Baum:2018PRL,Peng:2018PRB} & Floquet states & Temporal modulation & No upper limit \\ \hline
  \end{tabular}
  \label{summarytable}
\end{table*}

\subsection{Atomic states} In this original scheme of synthetic dimensions in atomic gases, a discrete set of internal atomic states is re-interpreted as a set of distinct lattices sites along an extra dimension~\cite{Boada:2012PRL, Celi:2014PRL}. These states are then coherently coupled by suitable external laser fields to simulate the `hopping' along the synthetic dimension. Importantly, these couplings can naturally be used to engineer topological models, as they include a spatially-dependent and tunable phase-factor $\propto \text{exp}(i {\bf q}\cdot {\bf r})$, which can mimic the Peierls phase-factor described before, and which physically reflects the net momentum transfer ${\bf q}$ from the lasers to the atoms in the transitions. It must be noted that the hopping amplitudes are inhomogeneous along the synthetic dimension, as coupling-matrix elements vary for different atomic transitions; however, this anisotropy is not expected to destroy features like topological edge modes in realistic systems~\cite{Celi:2014PRL}. 

\begin{figure}
%\resizebox{0.48\textwidth}{!}{\includegraphics{atompicture2.png}}
\resizebox{0.48\textwidth}{!}{\includegraphics{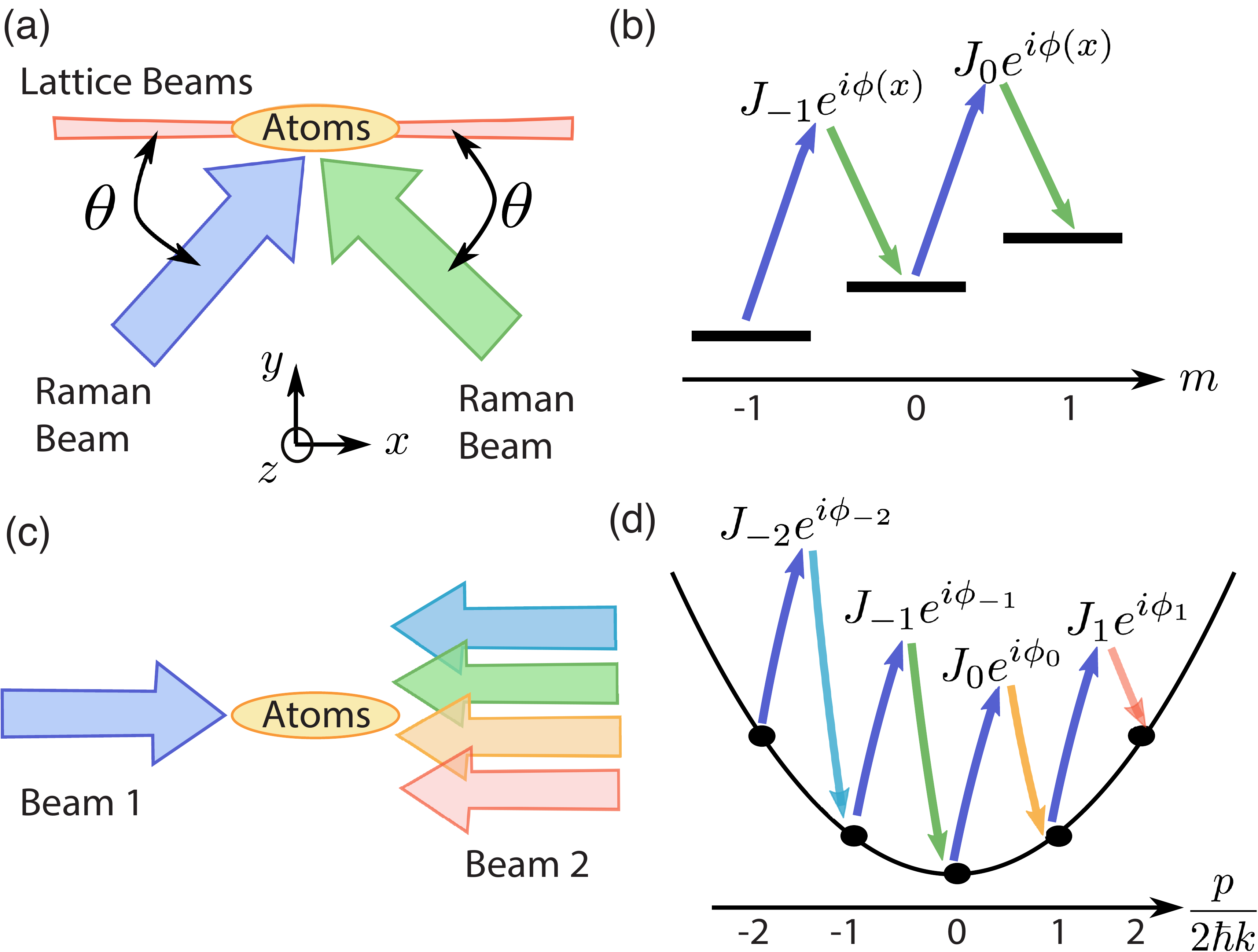}}
\caption{{\bf Main experimental set-ups used to realise synthetic dimension for atomic gases.} (a)\&(b) Schematic illustration of the atomic hyperfine-state scheme~\cite{Celi:2014PRL,Mancini:2015Science,Stuhl:2015Science} in which Raman lasers provide external couplings between a set of different Zeeman states, which lie within a given hyperfine manifold and are distinguished by a quantum number, $m$. The `hopping' in the synthetic dimension corresponds to a two-photon process, as indicated by arrows in (b), leading to a transition-dependent amplitude, $J_m$ from state $m$ to $m+1$, and a common complex phase, $\phi(x)$, set by the net momentum transfer from the lasers to the atoms. The latter can be designed to realise an artificial magnetic field (for example as in Eq.~\ref{eq:HH}) that is controlled via the angle $\theta$ shown in (a), between the Raman beams and the lattice beams confining the atoms to move along $x$. (c)\&(d)~Schematic illustration of a momentum-state scheme~\cite{Gadway:2015PRA,Meier:2016PRA} in which two counter-propagating lasers induce transitions between a set of discrete momentum states. One of the beams in (c) contains multiple frequency tones, so as to resonantly address each required two-photon transition, as shown in (d); this individual control means that all amplitudes,  $J_{p_n}$, and phases, $\phi_{p_n}$, along the synthetic dimension are fully tuneable. The scheme depicted here realises an effectively one-dimensional lattice; going further, additional lasers can implement a second synthetic dimension and, hence, engineer an artificial magnetic field~\cite{An:2017SciAdv}. 
} \label{fig:atompicture}
\end{figure}

Experimentally, the first implementation of this scheme presented a synthetic dimension with three atomic hyperfine states, coupled via Raman lasers~\cite{Mancini:2015Science,Stuhl:2015Science}. By adding the synthetic dimension to a 1D optical lattice, these set-ups effectively simulated a three-leg Hofstadter-like ladder, with a magnetic flux per unit cell set by the momentum transfer along the optical lattice direction, controlled via the angle between the Raman lasers beams and the lattice (Fig.~\ref{fig:atompicture}(a)\&(b)). The experiments took advantage of the sharp `edges' (boundaries) in the synthetic dimension and of powerful state-selective measurement techniques to observe chiral edge motion for the first time in ultracold atoms. Since these initial developments, synthetic dimensions have also been proposed and realised with pairs of long-lived electronic states~\cite{Wall:2016PRL,Livi:2016PRL,Kolkowitz:2017Nature} coupled via optical clock transitions, demonstrating the wide applicability of this approach.  

The restriction of recent experiments to short synthetic dimensions of two~\cite{Livi:2016PRL,Kolkowitz:2017Nature}, three~\cite{Mancini:2015Science,Stuhl:2015Science} or, recently, five~\cite{genkina2018imaging} sites reflects on-going experimental challenges in simultaneously coupling together many atomic states. Nevertheless, new research has shown that it is possible to extract a signature of the topological Chern number even in a narrow strip~\cite{genkina2018imaging,mugel2017measuring}, and to experimentally implement periodic boundary conditions in the synthetic dimension~\cite{han2018band,Li:arXiv2018}, reducing some finite-size effects. In the future, it may also be possible to expand the number of states up to the order of ten~\cite{Celi:2014PRL} by targeting atomic species with many suitable internal states, such as $^{40}K$. Finally, using atomic states as a synthetic dimension  may allow the realisation of new lattice topologies~\cite{Boada:2015NJP}, such as cylinders or M\"{o}bius strips, or of complicated lattice geometries, such as semi-synthetic zig-zag~\cite{Anisimovas:2016PRA,Xu:2018PRL}, triangular or hexagonal lattices~\cite{Suszalski:2016PRA}.  \\

\subsection{Momentum states} In physics, there is a powerful duality between position and momentum, such that a physical system can be equivalently described in either position space or momentum space. If the roles of position and momentum are reversed, this can lead to momentum-space topological physics, ranging from momentum-space lattices~\cite{Cooper:PRL2012, Ozawa:PRA2015}, to momentum-space Landau levels~\cite{Price:2014PRL,Berceanu:2016PRA}, and momentum-space integer~\cite{Ozawa:PRB2016} and fractional~\cite{Claassen:PRL2015} quantum Hall effects. In the context of synthetic dimensions, the position-momentum duality is reflected in a re-interpretation of a discrete set of free-particle momentum states as the set of lattice sites~\cite{Gadway:2015PRA,Meier:2016PRA,Viebahn:2019}. 

Experimentally, a set of momentum states can be selected by illuminating the atomic gas with two counter-propagating laser fields, far-detuned from any atomic transitions (Fig.~\ref{fig:atompicture}(c)\&(d)). Unlike the scheme using atomic states (Fig.~\ref{fig:atompicture}(a)\&(b)), the atoms therefore remain in the same atomic state, but scatter light by absorbing a photon from one beam and emitting into the other. For laser fields of wave-vector ${\bf k}$, momentum conservation means that such two-photon Bragg transitions changes the atomic momentum by $2 \hbar {\bf k}$. Starting with an atomic Bose-Einstein condensate at rest, this selection rule defines an infinite set of momenta, ${\bf p}_n \!= \!2 n \hbar {\bf k}$, where $n$ is an integer that can label sites along the synthetic dimension. Moreover, for free-particle states, the energy difference, $\!E_{n+1}\!-\! E_n \!\propto \!{\bf p}_{n+1}^2\! - \!{\bf p}_{n}^2$, varies between different pairs of states, such that each transition can be addressed separately by adding a suitable frequency to one of the laser fields. This leads to a key advantage of this scheme; as each transition is controlled separately, there can be arbitrary local and time-dependent control of all resulting parameters, including each effective `hopping' amplitude and phase~\cite{Gadway:2015PRA, Meier:2016PRA, an2017diffusive}. Experimentally, this has been used to explore interesting 1D topological phases -- such as the Su--Schrieffer--Heeger model~\cite{Meier:2016NatComm} and the topological Anderson insulator~\cite{Meier:2018Science}-- with time-of-flight measurements giving full access to the momentum distribution and, hence, to a `single-site' resolution in the synthetic space.

A second synthetic dimension has also been realised experimentally by adding a second pair of co-propagating beams~\cite{An:2017SciAdv} to couple states with: ${\bf p}_{n,m} \!= \!2 \hbar (n {\bf k}_1 +m {\bf k}_2) $, where ${\bf k}_{1}$ and ${\bf k}_2$ are the wave-vectors of the beam-pairs, and $(n,m)$ are then the effective site indices. The individual control of each `hopping' phase can be used to write-in a magnetic field for the 2D lattice; in practice, it is experimentally challenging to have many sites in both dimensions~\cite{Gadway:2015PRA,An:2017SciAdv}, but the proof-of-principle was demonstrated with a Harper-Hofstadter model of $5 \times 2$ sites~\cite{An:2017SciAdv}. Recently, another scheme~\cite{Cai:arxiv2018} has also measured chiral edge currents with room-temperature atoms in a momentum-space zig-zag ladder of timed Dicke states called a  superradiance lattice~\cite{Wang:2015PRL,Wang:2015Optica,Chen:2018PRL2}, bringing momentum-space synthetic dimensions to a new experimental platform.\\

\subsection{Other methods} As the above schemes have already been implemented, there have also been several theoretical proposals for other methods to engineer topological models with synthetic dimensions in atomic gases. 

Firstly, a synthetic dimension could be realised by shaking a harmonic trap~\cite{Price:2017PRA}. In this approach, the standard 1D harmonic oscillator states serve as lattice sites in the synthetic dimension, with a weakly anisotropic `hopping' induced by the time-dependent modulation. This could provide a very long synthetic dimension, limited only by the anharmonicity of the trap. When the system is extended along a real spatial dimension, it could, for example, realise a Hofstadter-like model, with a magnetic flux set by a spatially-varying modulation phase~\cite{Price:2017PRA}. This method can be used to directly measure the quantum Hall current in a one dimensional atomic wire coupled to reservoirs~\cite{Salerno:2018arXiv}.

Secondly, a synthetic dimension could be implemented by coupling the rotational states of ultracold polar molecules with microwaves~\cite{Sundar:2018SciRep,Sundar:2018arXiv}.This could again lead to a very long synthetic dimension, taking advantage of the rich spectrum of rotational states. Like the momentum-state scheme above, each transition would be driven separately, allowing for full control of `hopping' parameters and for the simulation of topological models. Another related approach could be to harness Rydberg states, as a large set of these can be experimentally coupled together with radio-frequency fields~\cite{Signoles:2017PRL}. 

Finally, a recent proposal for so-called `Floquet synthetic dimensions'~\cite{Martin:2017PRX} applies generally to systems under strong external drives. Floquet theory is used to model the effects of time-periodic modulations, showing that a quantum state is dressed by all harmonics of the driving frequency. In this proposal, the different harmonics are re-interpreted as different sites along a `spatial' dimension, where the `hopping' corresponds to absorbing or emitting a photon. When additional incommensurate driving frequencies are added, this picture leads to a multi-dimensional synthetic `Floquet lattice', which can be designed with topological properties~\cite{Martin:2017PRX,Baum:2018PRL,Peng:2018PRB,Andrijauskas:2018NJP}. \\

\section{Synthetic dimensions in photonics}

\noindent The other main platform in which synthetic dimensions have been actively studied is photonics ~\cite{Yuan:2018Optica}. In this field, the 
simulation of condensed matter phenomena is a rapidly-growing research direction~\cite{Carusotto:2013RMP}, with synthetic dimensions providing new routes towards realising topological lattice models and related possible applications. We briefly introduce various proposals and realisations for synthetic dimensions in photonics, again highlighting selected characteristics in Table 1. \\

\subsection{Orbital angular momentum modes} The first photonics proposal for a topological model with a synthetic dimension was based on the orbital angular momentum states of light in a cavity~\cite{Luo:2015NatComm}. In this proposal, a set of cavity modes with different orbital angular momenta, but the same resonant frequency,  are coupled together via spatial light modulators. By viewing the orbital angular momentum modes as a synthetic dimension, this approach was proposed for the realisation of the Harper-Hofstadter model (Eq.~\ref{eq:HH}) in a one-dimensional cavity array~\cite{Luo:2015NatComm}. Subsequently, this idea was pushed forward introducing sharp edges along this synthetic dimension~\cite{Zhou:2017PRL}, which is advantageous for investigating topological edge-state physics.
Orbital angular momentum states have also been used experimentally as a synthetic dimension in a quantum walk to realize one-dimensional~\cite{Cardano:2017NatComm} and two-dimensional~\cite{Wang:2018PRL} topological models.\\

\subsection{Frequency modes}
Another approach to realising a synthetic dimension in photonics is to use the frequency modes~\cite{Schwartz:2013OptExp} of light in a multi-mode cavity. Ring resonators, for example, can support multiple whispering-gallery modes, corresponding to resonances of light circulating around the cavity via total internal reflection. From the resonance condition for standing waves in a ring, it can be shown that the resulting modes are (approximately) equally-spaced in frequency. Such modes could then be coupled together into a synthetic dimension, by applying an appropriate time-dependent modulation of the cavity refractive index at a frequency equal to that of the mode frequency spacing. This  allows photons to be transferred between different modes~\cite{Ozawa:2016PRA,Yuan:2016OptLett}. Suitable Peierls `hopping' phases may be introduced in the synthetic direction by varying the modulation phase between different resonators. This proposal has also been extended to topological models of arbitrary dimension~\cite{Yuan:2018PRB}, by including appropriate extra modulation frequencies that add longer-range `hoppings' in the synthetic dimension which can be viewed as embedding a higher-dimensional lattice in one (synthetic) dimension~\cite{Schwartz:2013OptExp,Tsomokos:2010PRA,Jukic:2013PRA}.

Experimentally, a synthetic dimension of frequency modes has been recently realized for another system, consisting of light propagating in a nonlinear medium~\cite{Bell:2017Optica}. In this experiment, a weak signal beam propagates through a medium in the presence of at least two strong pump beams whose frequencies are shifted by an amount $\Omega$. Due to the pump beams, the signal frequency can change by $\pm \Omega$, mimicking time evolution in a one-dimensional non-topological tight-binding lattice of frequencies spaced by $\Omega$. If pump beams contain light whose frequency is shifted by $2\Omega$ from the other beams, the signal undergoes next- and next-next-nearest neighbor `hoppings', realizing models with long-range hoppings.\\

\subsection{Spatial eigenmodes} 
\begin{figure}
\resizebox{0.49 \textwidth}{!}{\includegraphics*{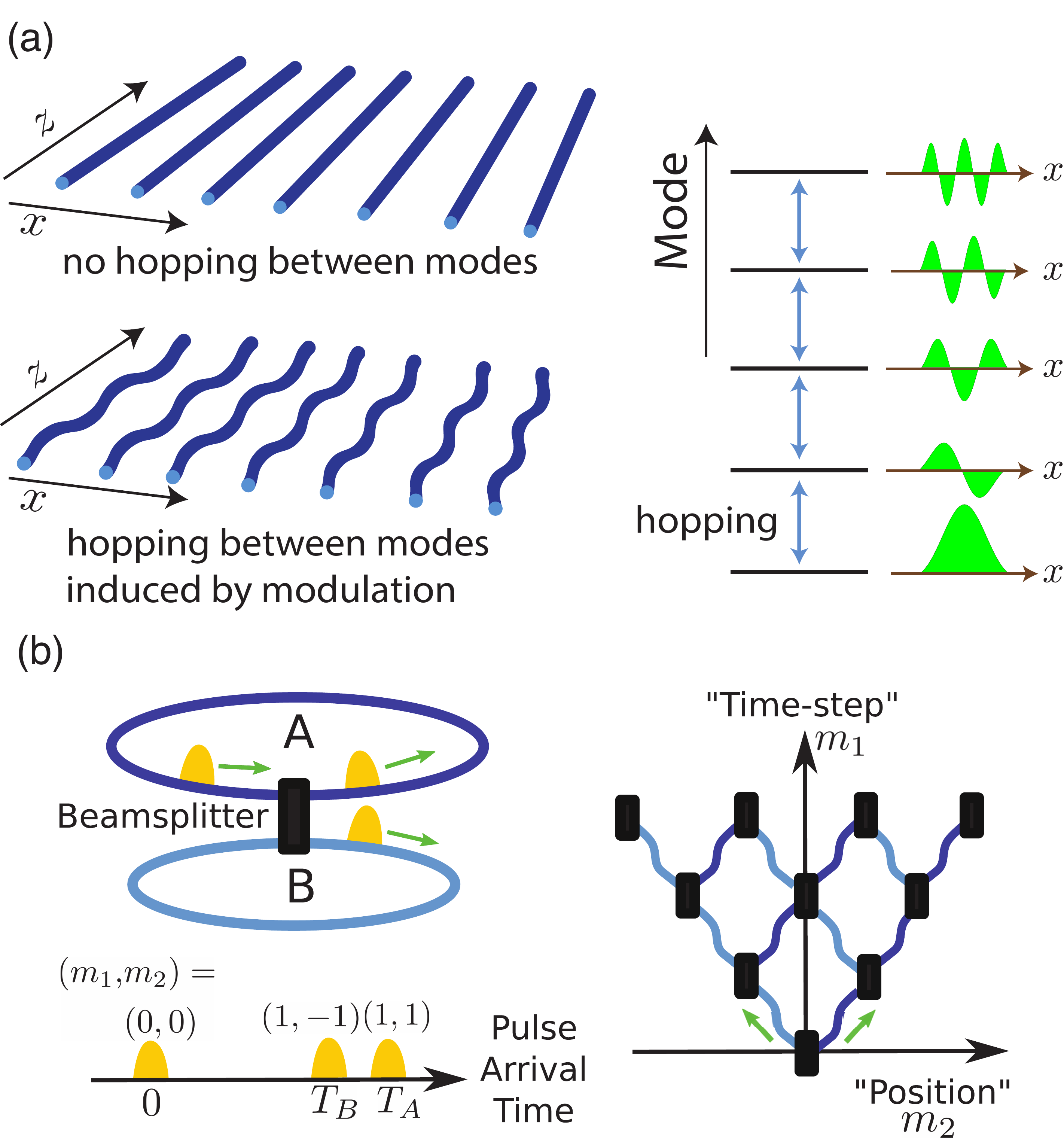}}
\caption{{\bf Different approaches towards synthetic dimensions in photonics}. (a) Synthetic dimension consisting of the spatial modes of a waveguide array~\cite{Lustig:2018arXiv}. In this system, light can propagate along the $z$ direction, while being in different spatial modes along the $x$ direction.
(Right) The structure of the array is carefully calibrated so that these modes along $x$ are equally spaced with respect to the propagation constant; schematic mode profiles are also illustrated. (Upper left) For straight waveguides along $z$, the modes are decoupled. (Lower left) By making the waveguides spatially oscillate along the $z$ direction, the spatial modes along $x$ can be coupled to realise a synthetic dimension.
Experimentally, this scheme was extended to implement a two-dimensional topological lattice, in which the topological edge states have been observed~\cite{Lustig:2018arXiv}.
(b) Schematic example of how a synthetic dimension can be encoded in the arrival time of pulses through a time-multiplexing approach~\cite{Schreiber:2010PRL,Regensburger:2011PRL}. Light pulses propagate through two long coupled fiber loops of slightly different lengths (upper left) leading to the observation of a train of pulses over time (lower left). As the light takes a time $T_A$ ($T_B$) to circulate around the $A$ ($B$) loop, the arrival time of the pulses at the observation point can be used to infer a pair of integers $(m_1, m_2)$, where $m_1$ counts the total number of round-trips in either loop, and $m_2$ counts how many more round trips were made in the long compared to the short loop. (Right) Interpreting $m_1$ as a `time-step' and $m_2$ as the `position' maps the light pulses to a 1D lattice at discrete times.} \label{photonicfigure}
\end{figure}

A topological model with a synthetic dimension has recently been experimentally realised by using different spatial modes of a waveguide array~\cite{Lustig:2018arXiv}. In such platforms, the paraxial propagation of light along a waveguide is described by a conservative evolution equation, which is analogous to the Schr\"{o}dinger equation, where the roles of time, $t$, and the propagation direction, $z$, swapped around~\cite{boyd2003nonlinear,ozawa2018topological}. To construct a synthetic dimension in such a system~\cite{Lustig:2018arXiv}, a one-dimensional array of $N$ waveguides was designed such that along $x$, the direction perpendicular to propagation, the array supported $N$ different spatial modes (Fig.~\ref{photonicfigure}(a)). These modes were calibrated to be equally spaced with respect to the propagation constant, which here plays the role of frequency. These equi-spaced levels were then coupled by spatially oscillating the waveguides along the propagation direction, $z$. In the analogy with the Schr\"{o}dinger equation, this corresponds to shaking the system in time, similar to how atomic harmonic trap states may be coupled by shaking the trap~\cite{Price:2017PRA}. Experimentally, multiple copies of these one-dimensional arrays were combined with different modulation phases to realise a Harper-Hofstadter model (Eq.~\ref{eq:HH}) with $N=7$ sites along the synthetic dimension. This comparatively large system size allowed to observe of edge-state propagation in a synthetic dimension: a hallmark of topological lattice models.\\

\subsection{Pulse arrival time}
Another class of synthetic dimensions uses time as a degree of freedom by employing time-multiplexing schemes~\cite{Schreiber:2010PRL,Regensburger:2011PRL}. To illustrate this concept, let us consider a system of two long optical fiber loops, which have a small length difference and which are coupled by a beamsplitter (Fig.~\ref{photonicfigure}(b)). A light pulse injected into one loop will be split by the coupler into two pulses: one circulating around the longer loop ($A$) and the other in the shorter loop ($B$). The pulses arrive back at the beamsplitter after a time-delay $T_A$ and $T_B$ respectively, before being again split by the beamsplitter, and so on. Monitoring the system at the beamsplitter, we observe a train of light pulses in time. 

Interestingly, the arrival time of each pulse can be re-interpreted to give a `position' index along a synthetic dimension; to see this, we introduce two time-scales: the average time-delay, $\bar{T}\!=\!(T_A+T_B)/2$, and the time-delay difference $\Delta T \!= \!T_A-T_B $. At early times, if there is a clear separation of these time-scales ($\bar{T}\! \gg  \!\Delta T$), it is possible to uniquely express the arrival time of a given pulse as $T \!=\! m_1 \bar{T} + m_2 \Delta T/2$, where the integer $m_1$ counts the total number of round-trips in either loop, and the integer $m_2$ counts how many more round trips were made in the long compared to the short loop. As light propagates, the integer $m_1$ increases to count each successive round-trip, while the integer $m_2$ can increase or decrease depending on which loop is traversed. This motivates us to re-interpret $m_1$ as a continually-increasing discrete `time-step' and $m_2$ as a discrete `position' index (Fig.~\ref{photonicfigure}(b)). Such time-multiplexing schemes have been exploited to simulate a wide-range of phenomena in synthetic 1D lattices, including quantum random walks~\cite{Schreiber:2010PRL,Schreiber:2011PRL,Regensburger:2011PRL,vatnik2017anderson}, the effects of PT-symmetry~\cite{Regensburger:2012Nature,Regensburger:2013PRL,Wimmer:2015NatComm} and geometrical pumping~\cite{Wimmer:2017NatPhys}. Incorporating different time scales for the arrival times, the time-multiplexing approach can also be extended to more dimensions~\cite{schreiber20122d}, as has also recently been experimentally used to explore a two-dimensional topological quantum walk~\cite{Chen:2018PRL}.\\

\subsection{Other methods}
Besides the schemes discussed above, there have been other theoretical proposals for synthetic dimensions in photonics and related areas. Indeed, one of the earliest proposals outside of cold atoms was for a synthetic dimension in optomechanics, consisting of two or three lattice sites made of photon and phonon degrees of freedom~\cite{Schmidt:2015Optics,Poshakinskiy:2017PRL}. Another proposal is closely related to the schemes using the frequency modes of a ring resonator~\cite{Ozawa:2016PRA,Yuan:2016OptLett}, with the angular coordinate within the ring being used as a synthetic dimension~\cite{Ozawa:2017PRL}. This scheme has unique features: the synthetic dimension is continuous rather than discrete, and  inter-particle interactions become local even along the synthetic dimension, as we discuss below.\\

\section{Prospects}

\noindent Synthetic dimensions not only provide alternative means of realizing previously-known topological phenomena, but they can also open doors to exploring physics that is not otherwise accessible. We conclude this Perspective by highlighting some unique features of synthetic dimensions and discussing future research prospects.\\

\subsection{Interparticle interactions and many-body physics} 

\noindent Perhaps the most important difference between real and synthetic dimensions, in the context of condensed matter physics, is the difference in inter-particle interactions. Usually in real materials, spatially-close particles interact stronger than those that are far apart, which means the inter-particle interaction has some degree of locality. Conversely, in many implementations of synthetic dimensions, particles that are far apart in the synthetic dimension interact as strongly as those that are close together. This is because particles that are far apart in synthetic space are still nearby in real space, and so can still strongly interact. It is also possible that interactions may not conserve the `position' along the synthetic direction: collisions, for example, can change atomic states~\cite{Chang:2005NatPhys}. From the synthetic dimension point-of-view, these exotic inter-particle interactions may have significant effects on many-body physics.

The most studied form of inter-particle interactions along a synthetic dimension is infinite-ranged with SU$(N$) symmetry, where $N$ is the number of lattice sites along the synthetic direction. The Hamiltonian of this interaction is~\cite{Celi:2014PRL}:
\begin{align}
	\hat{H}_\mathrm{int} &= g \sum_{\mathbf{r}} \mathcal{N}_\mathbf{r} \left(\mathcal{N}_\mathbf{r} - 1\right),
	&
	\mathcal{N}_\mathbf{r} &= \sum_{m} \hat{a}^\dagger_{\mathbf{r},m} \hat{a}_{\mathbf{r},m},
\end{align}
where $g$ is the interaction strength, and $\hat{a}_{\mathbf{r},m}$ is the annihilation operator of a particle at real position $\mathbf{r}$ and synthetic position $m$.
Such inter-particle interactions appear in the context of ultracold atomic gases with alkaline-earth atoms, where the synthetic dimension is spanned by nuclear spin states~\cite{Mancini:2015Science}. Such interactions do not distinguish different positions along the synthetic direction, making these very different from local interactions along real dimensions.

The effect of these interactions has been actively studied in the experimentally-relevant~\cite{Mancini:2015Science} ladder geometry, in which the real dimension is long and the synthetic dimension has only a few sites. For fermions, intriguing gapped states in the presence of magnetic flux with fractional filling and charge/spin order have been found~\cite{Barbarino:2015NatComm}. Additionally, edge physics~\cite{Yan:2015SciRep,Cornfeld:2015PRB,Barbarino:2016NJP}, fractional pumping~\cite{Zeng:2015PRL,Taddia:2017PRL}, Creutz-Hubbard models~\cite{Junemann:2017PRX} and exotic bound states~\cite{Ghosh:2015PRA,Ghosh:2017PRA} have also been studied. For bosons, charge density wave states, one-dimensional Haldane phases and other exotic phases have been identified~\cite{Anisimovas:2016PRA,Greschner:2016PRA,Greschner:2017PRL,Bilitewski:2016PRA,Xu:2018PRL}.

Going beyond ladder geometries to longer synthetic dimensions, a crucial open question with no consensus being reached as yet is whether this unique form of interactions can give rise to familiar two-dimensional phenomena, such as two-dimensional fractional quantum Hall phases~\cite{Calvanese:2017PRX,Lacki:2016PRA,Saito:2017PRA}. However, as mentioned above, a synthetic dimension based on the angular coordinate of a photonic ring resonator has been proposed, and in this scheme, local interactions in the resonator would also lead to local interactions along the synthetic dimension~\cite{Ozawa:2017PRL}. This may therefore open the way to observing the more familiar many-body physics of local interactions with synthetic dimensions.

Other implementations of synthetic dimensions can also have other forms of inter-particle interactions. For a synthetic direction made of harmonic oscillator eigenstates, the inter-particle interactions decay algebraically~\cite{Price:2017PRA}, whereas that made of frequency modes in photonics~\cite{Ozawa:2016PRA,Yuan:2016OptLett} preserves the total synthetic positions of colliding particles, that is, $\propto \sum_{m_1 + m_2 = m_3 + m_4} \hat{a}^\dagger_{m_1} \hat{a}^\dagger_{m_2} \hat{a}_{m_3} \hat{a}_{m_4}$. The momentum states of ultracold gases~\cite{Gadway:2015PRA} give rise to effective locally attractive interaction in momentum states, and the localization transition under such an interaction has been experimentally investigated~\cite{An:2018PRL,An:2018PRX}.
Theoretical works have also studied the effect of dipolar interactions when the synthetic dimension is spanned by the rotational states of polar molecules, and have found phase transitions to interesting states, localized along the synthetic direction~\cite{Sundar:2018SciRep,Sundar:2018arXiv}.
However, the implications of all these different scenarios and types of interactions on many-body phases of matter in synthetic lattices have been little understood; further investigation is much awaited.\\

\subsection{Applications} 
\begin{figure}
\resizebox{0.49 \textwidth}{!}{\includegraphics*{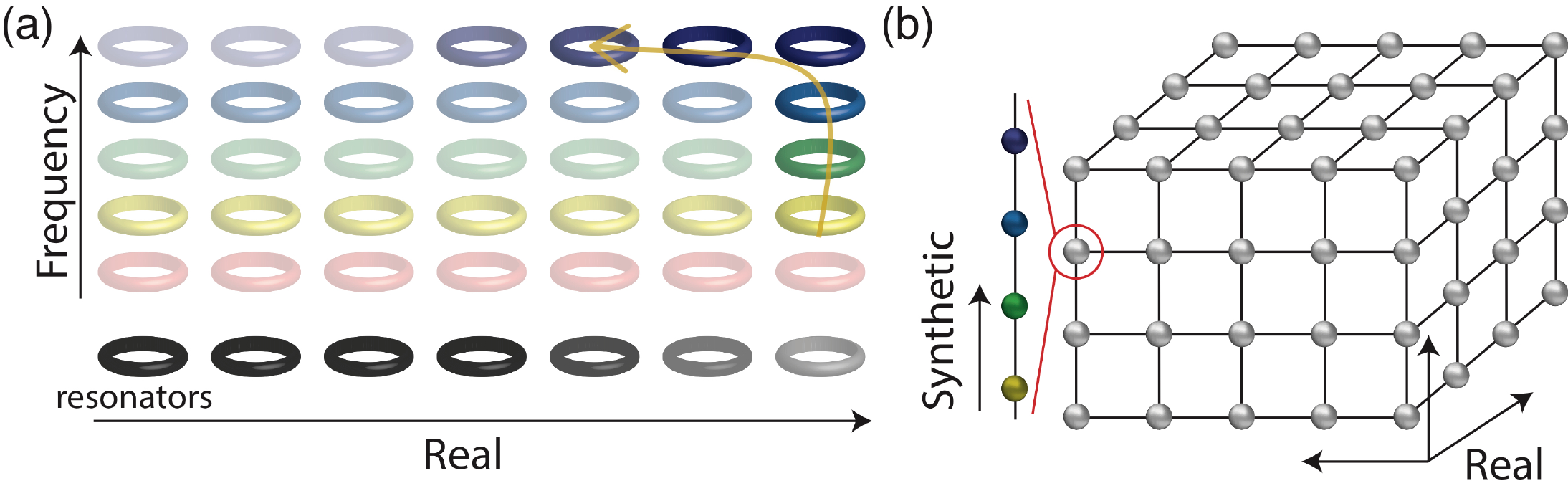}}
\caption{{\bf Prospects of synthetic dimensions} (a) Schematic illustration of a one-dimensional chain of optical resonators, simulating a two-dimensional topological lattice. The edge mode traveling along the frequency axis acts as a frequency converter, whereas the edge mode traveling along the real axis acts as an optical isolator. (b) Schematic image of a four-dimensional lattice made of three real dimensions and one synthetic dimension, which can be used to realize the four-dimensional quantum Hall effect.} \label{FigProspect}
\end{figure}

Besides providing a flexible platform to engineer various topological lattice models, synthetic dimensions may also offer means to realize devices which are otherwise difficult to fabricate.
In photonics, for example, there has been interest in using the chiral edge states of two-dimensional quantum Hall systems to build an optical isolator: a device which transmits light only in one direction. Without synthetic dimensions, one needs a bulk two-dimensional structure to obtain these edge states, leading to a device with a large footprint, which is a disadvantage for integrated photonic circuits. However, if one dimension can be made synthetic, chiral edge modes can be exploited in more compact, spatially one-dimensional structures~\cite{Ozawa:2016PRA}.  A schematic illustration of how the edge states in a synthetic two-dimensional lattice can be used as an optical isolator is presented in Fig.~\ref{FigProspect}(a).
Other potential applications of photonic synthetic dimensions include quantum memory and optical filters~\cite{Luo:2017NatComm}, as well as high-efficiency frequency conversion with photonic wavepackets propagating either along an edge in the frequency direction~\cite{Yuan:2016OptLett}(Fig.~\ref{FigProspect}(a)), or under periodic switching of a `synthetic force' in the synthetic direction~\cite{Yuan:2016Optica},  \\

\subsection{Topological models in higher dimensions} 
Whereas most studies of topological matter in synthetic dimensions have focused on two-dimensional models, this approach can also be used to access topological physics in three or more spatial dimensions. In photonics, especially, it can be challenging to fabricate photonic lattices in three dimensions. In this regard, synthetic dimensions offer a promising alternative route to explore three-dimensional topological physics, such as Weyl points~\cite{Sun:2017PRA, Lin:2016NatComm} and weak topological insulators~\cite{Lin:2018SciAdv}.

Synthetic dimensions also open up the study of topological phenomena in dimensions of four or higher.
 For example, synthetic four-dimensional lattice can be constructed using three real dimensions and one synthetic dimension (Fig.~\ref{FigProspect}(b)).
From the celebrated periodic table of topological phases for non-interacting fermions~\cite{Kitaev:2009,Ryu:2010NJP,Chiu:2016RMP}, it is known that there should be a variety of topological phases above three dimensions. A first example of this is the four-dimensional quantum Hall effect~\cite{Frohlich:2000,Zhang:2001Science,Qi:2008PRB}, corresponding to a generalisation of the famous two-dimensional quantum Hall effect. For the simplest case of a non-degenerate band, the four-dimensional quantum Hall band structure is topologically characterized by an integer invariant called the second Chern number, $\mathcal{C}_2$, as opposed to the (first) Chern number, $\mathcal{C}_1$, for the two-dimensional quantum Hall systems, mentioned above. The first Chern number $\mathcal{C}_1$ leads to a Hall current that depends linearly on the applied electric field: $j^\mu = (e^2/2h)\mathcal{C}_1 \epsilon^{\mu \nu}E_\nu$, where $\epsilon^{\mu \nu}$ is the two-dimensional Levi-Civita symbol, which is totally anti-symmetric with respect to its indices. On the contrary, the second Chern number $\mathcal{C}_2$ leads to a nonlinear response $j^\mu = (e^3/2h^2)\mathcal{C}_2 \epsilon^{\mu\nu\rho\sigma}E_\nu B_{\rho\sigma}$, where $\epsilon^{\mu\nu\rho\sigma}$ is the four-dimensional Levi-Civita symbol. Here, $E_\nu = \partial_0 A_\nu - \partial_\nu A_0$ and $B_{\rho\sigma} = \partial_\rho A_\sigma - \partial_\sigma A_\rho$ are the four-dimensional electromagnetic fields applied to the system, with $\mathbf{A}$ being the four-dimensional electromagnetic gauge potential~\cite{Price:2015PRL}. Very recently, the first signatures of this four-dimensional quantum Hall effect have been experimentally observed through a mathematical mapping to two-dimensional topological pumps in ultracold atomic gases~\cite{Lohse:2018Nature} and photonics~\cite{Zilberberg:2018Nature}. Going further, synthetic dimensions can provide means to realise a fully four-dimensional physical set-up and so to directly observe the four-dimensional quantum Hall response in both cold atoms~\cite{Price:2015PRL,Price:2016PRB} and in photonics~\cite{Ozawa:2016PRA}. Finally, these developments have also stimulated studies of higher and more exotic topological states such as five-dimensional Weyl semimetals~\cite{Lian:2016PRB} and six-dimensional quantum Hall effects~\cite{Lee:2018PRB,Petrides:2018PRB}.\\

\section{Conclusions} Synthetic dimensions provide a powerful way to explore topological matter in cold atoms and photonics, opening up many interesting prospects for future research. As summarized in Table 1, the idea of synthetic dimensions has already led to a wealth of proposals and experiments, which each have their own distinct characteristics. An important research direction to take will be to understand the advantages and limitations of each set-up, and to work on their improvement. For example, most of the known schemes have exotic inter-particle interactions and only discrete degrees of freedom along the synthetic direction, which hinders the simulation of models with local interactions or with continuous dimensions; in the future, it will be interesting to find new approaches that circumvent or exploit these features. In this Perspective we have also focused on synthetic dimensions in atomic, molecular and optical systems, but similar physics may arise in other contexts, such as, for example, in superconducting phase qudits, where the $d$ anharmonic oscillator levels can be controllably coupled together~\cite{Neeley:Science2009}. Going forward, it will be important to develop and explore such broader connections, as the idea of topological matter in synthetic dimensions is very general and the extension of this approach to other areas of physics is much awaited. 

{\bf{Acknowledgements}}
T.O. was supported by JSPS KAKENHI Grant Number JP18H05857, RIKEN Incentive Research Project, and the Interdisciplinary Theoretical and Mathematical Sciences Program (iTHEMS) at RIKEN. H.M.P. was supported by funding from the Royal Society.

%{\bf{Author Contributions}}
%\\All authors have read, discussed and contributed to the writing of the manuscript. 

%{\bf{Conflict of Interest}}
%\\The authors declare no competing interests.

%{\bf{Publisher's note}}
%\\Springer Nature remains neutral with regard to jurisdictional claims in published maps and institutional affiliations.

\bibliographystyle{naturemag}
\bibliography{Bibliography_NatRevPhys}

\end{document}